\documentstyle[a4p,12pt,oldlfont,epsfig]{article}
\def\Journal#1#2#3#4{{#1}{\bf#2} (#4) #3}

\def\NIMA{{\sl Nucl. Inst. Meth.} \bf A}
\def\NPB{{\sl Nucl. Phys.} \bf B}
\def\PLB{{\sl Phys. Lett.} \bf B}
\def\PRL{\sl Phys. Rev. Lett. }
\def\PRD{{\sl Phys. Rev.} \bf D}
\def\ZPC{{\sl Z. Phys.} \bf C}
\newcommand{\etal} {{\it et al.,} }
\newcommand{\Zz}{\mbox{${\mathrm{Z}^0}$}}
\newcommand{\qq}{\mbox{$\mathrm{q\overline{q}}$}}
\newcommand{\Zqq}{\mbox{$\Zz/\gamma^*\rightarrow\qq$}}
\newcommand{\Jetset}{\mbox{JETSET}}
\newcommand{\Pythia}{\mbox{PYTHIA}}
\newcommand{\Herwig}{\mbox{HERWIG}}
\newcommand{\Phojet}{\mbox{PHOJET}}

\newcommand {\sqrts}         {$\sqrt{s}$}
\newcommand {\sqrtsp}        {\mbox{$\sqrt{s'}$}}

\newcommand{\Zg}             {\mbox{Z$^{0}\gamma$}}

\newcommand{\m}              {\mbox{$M$}}
\newcommand{\Dm}             {\mbox{$\Delta M$}}
\newcommand{\mX}             {\mbox{$M_{\rm X}$}}
\newcommand{\mY}             {\mbox{$M_{\rm Y}$}}
\newcommand {\ee}            {${\mathrm e}^+{\mathrm e}^-$}
\newcommand {\ra}            {\mbox{$\rightarrow$}}
\newcommand {\WW}            {${\mathrm W}^+ {\mathrm W}^-$}
\newcommand {\W}             {${\mathrm W}^{\pm}$}

\begin{document}
\begin{titlepage}
 \begin{center}{\Large   EUROPEAN LABORATORY FOR PARTICLE PHYSICS
 }\end{center}\bigskip
\begin{flushright}
  CERN-EP/98-013 \\  29 January 1998
\end{flushright}
\bigskip
%
\bigskip\bigskip
\begin{center}
{\huge \bf
Search for an Excess in the} \\
\vspace{3mm}
{\huge \bf  Production of
 Four-Jet Events from} \\
\vspace{1mm}
{\huge \bf {\boldmath $\mathrm{e}^+\mathrm{e}^-$\unboldmath} Collisions at
  {\boldmath$\sqrt{s} = 130-184$\unboldmath}~GeV}
\end{center}\vspace{0.5cm}

\begin{center}{\Large The OPAL Collaboration}
\end{center}\vspace{1cm}
\begin{center}{\large \bf Abstract}\end{center}
Events with four distinct jets from $\mathrm{e}^+\mathrm{e}^-$ collisions,
collected by the OPAL detector at centre-of-mass energies
between 130 and 184~GeV, are analysed for a peak in the sum of dijet masses.
This search is motivated by the ALEPH 
Collaboration's observation 
of a clear
excess of events with dijet mass sums close to 105 GeV
in data taken at  
centre-of-mass energies of 130 and 136 GeV in 1995.
We have observed no significant excess of four-jet events 
compared to the Standard Model expectation for any 
dijet mass sum at any energy. Our observation is inconsistent
with the excess observed by ALEPH in 1995.
Upper limits are determined on the production cross-section
as a function of the dijet mass sum.

\vspace{0.75cm}
\begin{center}{
Submitted to {\sl Physics Letters}
}
\end{center}


\end{titlepage}

\begin{center}{\Large        The OPAL Collaboration
}\end{center}\bigskip
\begin{center}{
K.\thinspace Ackerstaff$^{  8}$,
G.\thinspace Alexander$^{ 23}$,
J.\thinspace Allison$^{ 16}$,
N.\thinspace Altekamp$^{  5}$,
K.J.\thinspace Anderson$^{  9}$,
S.\thinspace Anderson$^{ 12}$,
S.\thinspace Arcelli$^{  2}$,
S.\thinspace Asai$^{ 24}$,
S.F.\thinspace Ashby$^{  1}$,
D.\thinspace Axen$^{ 29}$
% ...
}\end{center}\bigskip
\bigskip
$^{  1}$School of Physics and Astronomy, University of Birmingham,
Birmingham B15 2TT, UK
\newline
$^{  2}$Dipartimento di Fisica dell' Universit\`a di Bologna and INFN,
I-40126 Bologna, Italy
\newline
% ...
$^{  5}$Cavendish Laboratory, Cambridge CB3 0HE, UK
\newline
% ...
$^{  8}$CERN, European Organisation for Particle Physics,
CH-1211 Geneva 23, Switzerland
\newline
$^{  9}$Enrico Fermi Institute and Department of Physics,
University of Chicago, Chicago IL 60637, USA
\newline
% ...
$^{ 12}$Indiana University, Department of Physics,
Swain Hall West 117, Bloomington IN 47405, USA
\newline
% ...
$^{ 16}$Department of Physics, Schuster Laboratory, The University,
Manchester M13 9PL, UK
\newline
% ...
$^{ 23}$Department of Physics and Astronomy, Tel Aviv University,
Tel Aviv 69978, Israel
\newline
$^{ 24}$International Centre for Elementary Particle Physics and
Department of Physics, University of Tokyo, Tokyo 113, and
Kobe University, Kobe 657, Japan
\newline
% ...
$^{ 29}$University of British Columbia, Department of Physics,
Vancouver BC V6T 1Z1, Canada
\newline
% ...
University, Debrecen, Hungary
\newline
\newpage

\section{Introduction}

In a run of LEP in 1995 at centre-of-mass energies
of $\sqrt{s} = 130$ and 136~GeV, the ALEPH Collaboration 
observed~\cite{bib-ALEPH4jet} 
an excess of events with four distinct
jets compared with the Standard Model expectation.
Such an excess could be due to 
the production of new particles X and Y,
each decaying into two hadronic jets in the process
$\mathrm{e^+e^-} \ra \mathrm{XY} \ra$~four jets. 
The two particles could have equal or unequal masses.
Grouping the jets into pairs, calculating their pair 
invariant masses $M_{ij}$ and $M_{kl}$,
and selecting the combination yielding the smallest
mass difference $\Delta M \ = |M_{ij}-M_{kl}|$,
ALEPH observed a clustering of nine events in a mass window
6.3~GeV wide centred around $M = M_{ij}+M_{kl} \approx 105$~GeV,
with a Standard Model expectation of 0.8 events
in this mass window.
The choice of the combination with the minimum
$\Delta M$ would tend to favour the selection of particles
of equal mass or with a small mass difference.

In response to this observation, the OPAL Collaboration
performed an analysis that closely followed the
selection of Reference~\cite{bib-ALEPH4jet}.
In its 130 and 136~GeV data from 1995, OPAL
observed seven events with $M$ 
between 60 and 130~GeV, with an expected 
Standard Model background of $6.4 \pm 0.6$ events.
In the signal region indicated by ALEPH, OPAL observed
one event, consistent with the expected Standard Model background of 
$0.8 \pm 0.2$
events~\cite{bib-O135}.
The estimated efficiency of the OPAL analysis and the resolution on the
dijet mass sum are similar to those obtained by ALEPH.
Consequently, the OPAL detector would be expected to have a sensitivity
comparable to the ALEPH detector for a four-jet signal should one exist.
The ALEPH Collaboration also reported a slight excess
at the higher centre-of-mass energies of
161 and 172~GeV~\cite{jetstory}.
The DELPHI Collaboration
observed no significant peak at 105~GeV in a similar analysis using their
1995 data at 130 and 136 GeV~\cite{bib-D135}. 
The L3 Collaboration
likewise reported no excess of events in the indicated mass window
for $\sqrt{s} = 130$--172 GeV~\cite{bib-L135}.
Nonetheless,
there has been a great deal of theoretical speculation on the
cause of the excess observed by 
ALEPH~\cite{theory133a,theory133b,theoryhigh}.

In 1997, LEP made short runs at
$\sqrt{s} = 130$ and 136~GeV with an integrated luminosity
similar to that of 1995
at these centre-of-mass energies 
to test again the signal hypothesis.
We add these data to our sample described in Reference~\cite{bib-O135}, and 
also include data collected at 161, 172 and 183~GeV.
To search for the class of events observed by ALEPH
in a model independent fashion, we have performed analyses on the
OPAL data as close as possible to the ALEPH analyses at these
energies~\cite{jetstory}.  However, above the kinematic threshold for W-pair
production, a veto is imposed to suppress this new source of background and
results are presented with and without this requirement.

The comparison of results of the OPAL emulation of the
ALEPH selection to the ALEPH observation of an excess does not 
depend upon the underlying model of possible new physics if
only the number of observed events is compared.
In the context of a model of the process
$\mathrm{e^+e^-} \ra \mathrm{XY} \ra$~four jets,
a separate analysis
is also presented that is intended to improve the sensitivity for 
values up to 30~GeV of the difference in mass between the two produced
particles.  This broader search is motivated by the fact that
an analysis performed by the ALEPH Collaboration, 
using a kinematic fit which constrains the
masses of the two dijet systems to be equal~\cite{bib-ALEPHchargedhiggs},
indicates that the excess events are not consistent with the hypothesis
that the produced particles have equal mass.
Furthermore, compared with the ALEPH emulation analysis,
this OPAL-specific analysis is estimated to be more
sensitive to a four-jet signal of equal-mass particle production
at higher energies, and
its efficiency is less dependent on the flavour of the final-state 
quarks.
It is used in addition to the emulation of the
ALEPH analysis to set cross-section limits as function of
the dijet mass sum, and also to provide limits in the case of nonzero
mass difference.

\section{The OPAL Detector}

A detailed description of the OPAL detector can be found 
elsewhere~\cite{bib-OPALdet}.  
OPAL's nearly complete solid angle
coverage and excellent hermeticity enable it to detect
the four-jet final state with high efficiency.
The central tracking detector consists of a two-layer silicon
microstrip detector~\cite{SI} with polar 
angle\footnote{OPAL uses a right-handed coordinate system where the 
$+z$ direction is along the electron beam and where
$+x$ points to the centre of the LEP ring.  
The polar angle, $\theta$, is defined with respect to the 
$+z$ direction  and the azimuthal angle, $\phi$, with respect to
the $+x$ direction.} 
coverage $|\cos\theta| < 0.9$, immediately surrounding
the beam-pipe, followed by a high-precision vertex drift chamber,
a large-volume jet chamber and $z$-chambers, all in a uniform
0.435~T axial magnetic field.  A lead-glass electromagnetic calorimeter
is located outside the magnet coil, which, in combination with the 
forward calorimeter, gamma catcher and silicon-tungsten 
luminometer~\cite{SW}, complete the geometrical acceptance down 
to 24~mrad from the beam direction.  The silicon-tungsten luminometer
serves to measure the integrated luminosity using small-angle
Bhabha scattering events~\cite{lumitot}.
The magnet return yoke is instrumented with streamer tubes for hadron
calorimetry and is surrounded by several layers of muon chambers.

\section{Data and Monte Carlo simulations}

The data used in these analyses were collected
in five separate running periods.  The 
energies~\cite{energy} and integrated
luminosities~\cite{lumitot} 
for the five data samples are given in Table~\ref{table:elum}.
The 130 and 136~GeV data of 1995 and 1997 are collectively
referred to in this letter as the 133~GeV data.
The other three samples are referred to as 
the 161~GeV data, the 172~GeV data, and the 183~GeV data, and are analysed 
separately.

\begin{table}[tbh]
\begin{center}
\begin{tabular}{|c|c|c|}\hline
$\sqrt{s}$ (GeV) & Year & $\int\!{\cal L}{\rm d}t$ (pb$^{-1}$)
                        \\ \hline\hline
$130.3 $ & 1995 & 2.7  \\ 
$136.2 $ & 1995 & 2.6 \\ \hline\hline
$130.0 $ & 1997 & 2.6  \\ 
$136.0 $ & 1997 & 3.4  \\ \hline\hline
$161.3 $ & 1996 & 10.0  \\ \hline\hline
$172.1 $ & 1996 & 10.3 \\ \hline\hline
$182.7 $ & 1997 & 57.1 \\ \hline
\end{tabular}
\caption{{\sl Summary of the data samples, 
luminosity-weighted
centre-of-mass energies,  year of collection,
and integrated luminosities used in these analyses.}}
\label{table:elum}
\end{center}
\end{table}


The main backgrounds for the selection of anomalous four-jet events
are \Zqq\ production and Standard Model four-fermion production processes.
Monte Carlo samples modelling the backgrounds have been prepared
using \Pythia~5.7~\cite{pythia} for
the \Zqq\ process and 
EXCALIBUR~\cite{excalibur} and
grc4f~\cite{grc4f} for the Standard Model four-fermion processes,
all using \Jetset~7.4's parton shower and hadronization models~\cite{pythia}.
For the generation of Standard Model four-fermion processes, the W mass is taken
to be 80.33 GeV.  
Two-photon processes generated by \Pythia, \Herwig~\cite{herwig},
and \Phojet~\cite{phojet}
were used to estimate the contribution of these processes 
to the Standard Model background in the early stages of the analysis.

The signal detection efficiencies were estimated using
the HZHA generator~\cite{hzha} to simulate the production of
supersymmetric Higgs bosons
 $\mathrm{e}^+ \mathrm{e}^- \rightarrow 
\mathrm{h}^0 \mathrm{A}^0 \rightarrow
\mathrm{b\bar{b}b\bar{b}}$ as a model
for the signal process 
$\mathrm{e}^+ \mathrm{e}^- \rightarrow 
\mathrm{XY} \rightarrow 4$ jets.  Samples with decays into 
other quark flavours were also used to check for flavour dependence.
All Monte Carlo samples were processed through a full simulation of the
OPAL detector~\cite{bib-GOPAL}.

\section{Analysis and Results}

The main features of the signal process are four well-defined, energetic,
hadronic jets and a total visible event energy close to the
centre-of-mass energy.  
The Standard Model background expectation
changes considerably in size and composition
between $\sqrt{s}=$133~GeV and 183~GeV.  At 133~GeV,
the main background comes from 
$\mathrm{Z}^0/\gamma^* \rightarrow \mathrm{q\bar{q}}$
both with or without initial-state radiation and accompanied by
hard gluon emission. 
Above the threshold for
$\mathrm{e}^+\mathrm{e}^- \rightarrow \mathrm{W}^+\mathrm{W}^-$
at $\sqrt{s}=161$~GeV, the background from Standard Model
four-fermion processes is
important and becomes larger with increasing $\sqrt{s}$.
A procedure to reject 
$\mathrm{W}^+\mathrm{W}^-\rightarrow\mathrm{q\overline q q\overline q}$
is implemented for centre-of-mass energies of 161~GeV and above.

Events are reconstructed from charged particle tracks and energy
deposits (``clusters") in the electromagnetic and hadronic 
calorimeters.
Tracks are required to originate from close to the interaction point, 
to have more than a minimum number of hits in the jet 
chamber, and to have 
a transverse momentum greater than 0.1~GeV and a total momentum
less than 100~GeV~\cite{bib-hadsel161}.  
Energy clusters in the electromagnetic and hadron calorimeters 
are required to exceed minimum energy thresholds.
Tracks and clusters passing these quality requirements are then processed
to reduce double-counting of energy and momentum in the event by
matching charged tracks with calorimeter clusters.
The energy-momentum flow obtained with this algorithm~\cite{susypaper}
is used throughout the analysis.  
Energy measured in the forward detectors, covering
$|\cos\theta|>0.985$, has not been included
in the analyses presented here.

Selection criteria emulating the ALEPH analysis~\cite{bib-ALEPH4jet}
will first be described followed by the description of another set
of requirements for event selection in an OPAL-specific analysis.
Efficiencies and backgrounds for the two analyses are given followed
by estimates of systematic errors on these quantities.


\subsection{OPAL Emulation of the ALEPH Selection}
\label{secalephemul}

The procedure for selecting four-jet hadronic events and reconstructing
the dijet masses is described below, emulating the 
ALEPH selection as described in Reference~\cite{bib-ALEPH4jet} 
and subsequent modifications and
developments as described in Reference~\cite{jetstory}.
The number of events retained after each cut in sequence is given in
Table~\ref{table:datamc},
together with the expectation from \Zqq,
Standard Model four-fermion and two-photon background processes, for 
$\sqrt{s}=133$~GeV and for the sum over all
centre-of-mass energies.
Table~\ref{table:datamc} also lists the efficiencies
for the reference h$^0$A$^0$ Monte Carlo 
at $\sqrt{s}=133$~GeV after each step.


\begin{enumerate}

\item 
      Events are required to have at least five charged tracks
      and seven electromagnetic calorimeter clusters.
      The sum of the electromagnetic calorimeter
      cluster energies should be at least
      10\% of the centre-of-mass energy, and the electromagnetic calorimeter
      energy
      is required to be roughly balanced along the beam direction:
      $\sum E_i\cos\theta_i \leq 0.65\sum E_i$, where the sums run over
      measured electromagnetic calorimeter
      clusters.  The properties of this selection are
      detailed in Reference~\cite{bib-hadsel161}.

\item 
      To remove events with a real Z$^0$ and
      large initial-state radiation (radiative return events), 
      events must satisfy 
      $|p_z^{\mathrm{vis}}|\ \leq \
       K(M_{\mathrm{vis}}-90 \,\,\mathrm{(GeV)})$,
      where $p_z^{\mathrm{vis}}$ is the momentum sum
      along the beam direction and
      $M_{\mathrm{vis}} = \sqrt 
      { E_{\mathrm{vis}}^2 - p_{\mathrm{vis}}^2  }$ is the
      total observed mass. $K$ is a coefficient depending on
      $\sqrt{s}$. For the 133~GeV sample, $K=0.75$; for
      the 161~GeV sample, $K=1.50$; and for the
      172~GeV and 183~GeV samples, $K=1.65$.

\item 
      Jets are formed with the
      Durham jet-finding algorithm~\cite{bib:durham}
      with its resolution parameter $y_{\mathrm{cut}}$ set to 0.008.
      Selected events are required to have four or more jets.
      For events with five or more jets, 
      the jet pair with the smallest invariant
      mass is combined into a single jet and this procedure is
      repeated until four jets are left.  

\item 
      The contribution of radiative
      return events is further reduced by requiring for each jet that
      the energy observed in the electromagnetic calorimeter, after
      subtracting the energy expected to have been deposited by the
      jet's charged hadrons, is less than 80\% of the jet energy.
      The expected hadronic energy in the electromagnetic calorimeter
      is calculated using a 
      track-cluster matching algorithm~\cite{susypaper}.

\item 
      All jet masses are required to exceed 1.0 GeV to suppress further
      the contribution from radiative return events.

\item 
      The energies and momenta of each jet are rescaled imposing 
      conservation of energy and momentum for the event using the beam
      energy constraint.  
      The jet velocities $\vec\beta_i=\vec p_i/E_i$ are held
      fixed in the scaling.
      If one or more scaling factors is negative, the event is
      rejected.  The rescaled jet energies and momenta are used
      in the following stages of the selection.


\item 
      To suppress events involving gluon radiation,
      all two-jet combinations are required to have an invariant
      mass of more than 19.2\% of the centre-of-mass energy.  This
      initially corresponded to 25 GeV for the 133~GeV sample.

\item 
       All combinations of jet pairs must 
       have a sum of the individual jet masses $(M_i+M_j) > 10$~GeV.

\item 
      All combinations of jet pairs must have a total charged
      multiplicity of at least 10.

\item W-pair veto:
      For the 161 GeV data sample, a requirement is placed on the
      dijet mass sum for each of the three possible pairings of jets:
      $M \leq 150$~GeV for the pairing with the smallest $\Delta M$,
      $M \leq 152$~GeV for the pairing with intermediate $\Delta M$, and
      $M \leq 156$~GeV for the pairing with the largest $\Delta M$.
      For the 172 and 183~GeV data, it is required that
      $|M-160|\geq 10$~GeV for the pairing with the 
      smallest mass difference if $\Delta M$ 
      is less than 15~GeV, and the same condition is applied to the
      pairing with the second-smallest mass difference if 
      the second smallest $\Delta M$
      is less than 30~GeV.  No W-pair veto is
      applied to the 133~GeV sample.

\end{enumerate}

\begin{table}[tbhp]
\begin{small}
\vspace{-5mm}
\begin{center}
\begin{tabular}{|c||r|r|r|r|r|r|r|r|r|r|} 
\hline
Cut  & (1)    & (2)   & (3)  & (4)  & (5)  & (6)  & (7)  & (8) & (9) & (10) 
\\ \hline \hline
{\bf {\boldmath $\sqrt{s}=133$\unboldmath}~GeV}&& 
    &       &     &      &      &      &     &     &    \\
\cline{1-1}
\Zqq & 3151. & 1185. & 153.2 & 55.1 & 50.8 & 45.7 & 22.5 &18.0 &14.6 & 14.6 \\
4-Fermion&24.5& 16.1 &   4.2 &  3.6 &  3.1 &  3.0 &  1.6 & 1.1 & 0.7 & 0.7  \\
$\gamma
\gamma$&161.0 &  8.9 &   1.7 &  1.7 &  1.7 &  1.7 &  1.1 & 0.0 & 0.0 & 0.0  \\
\hline 
Total SM backg. &3337.& 1211.& 159.2 & 60.4 & 55.6 & 50.4 & 25.2 &19.1 &15.3 &
15.3  \\
\hline
Observed& 3372 &1165  &147   & 51   & 47   & 42   & 19   & 16  &13   & 13  \\
\hline 
Sig. effic., $\epsilon_{\mathrm{hA}}$ (\%) &
        99.8 & 90.2 & 69.0 & 68.6 & 67.4 & 67.2 & 65.4 & 48.6 & 45.4 & 45.4 \\
\hline
\hline
{\bf All {\boldmath $\sqrt{s}$ \unboldmath}}&&      
  &       &     &      &      &      &     &     &      \\
{\bf 130--184~GeV}&&      
  &       &     &      &      &      &     &     &      \\
\cline{1-1}
\Zqq & 10456. & 4450. & 481.1 & 188.7 & 176.7 & 
       155.0 & 68.2 & 59.7 & 49.2 & 40.2 \\
4-Fermion & 1098. & 913.0 & 476.0 & 428.0 & 400.5 & 
            364.8 & 258.1 & 236.3 & 191.7 & 74.2 \\
$\gamma
\gamma$& 450.7  & 26.3 & 2.7 & 2.2 & 2.1 & 1.8 & 1.0 & 0.0 & 0.0 & 0.0 \\
\hline
Total SM backg. &
 12004. & 5389. & 959.9 & 618.9 & 579.4 & 521.6 & 327.4 & 295.9 & 240.8 &
 114.3 \\
\hline
Observed & 12617 & 5461 & 984 & 635 & 592 & 527 & 328 & 299 & 240 & 92 \\
\hline
\end{tabular} 
\caption{{\sl Event counts observed by OPAL at the various
selection stages, with backgrounds estimated
using \Pythia\ for \Zqq, EXCALIBUR and grc4f for Standard Model four-fermion processes,
and \Pythia\ and \Phojet\ for two-photon processes.  
Signal efficiencies at 133~GeV for h$^0$A$^0$ (see text) 
for $M_{\mathrm h^0}=M_{\mathrm A^0}=55$~GeV are also listed.}}
\label{table:datamc}
\end{center}
\end{small}
\end{table}

\begin{figure}[htb]
\begin{center}
\epsfxsize=15.5cm
\epsffile{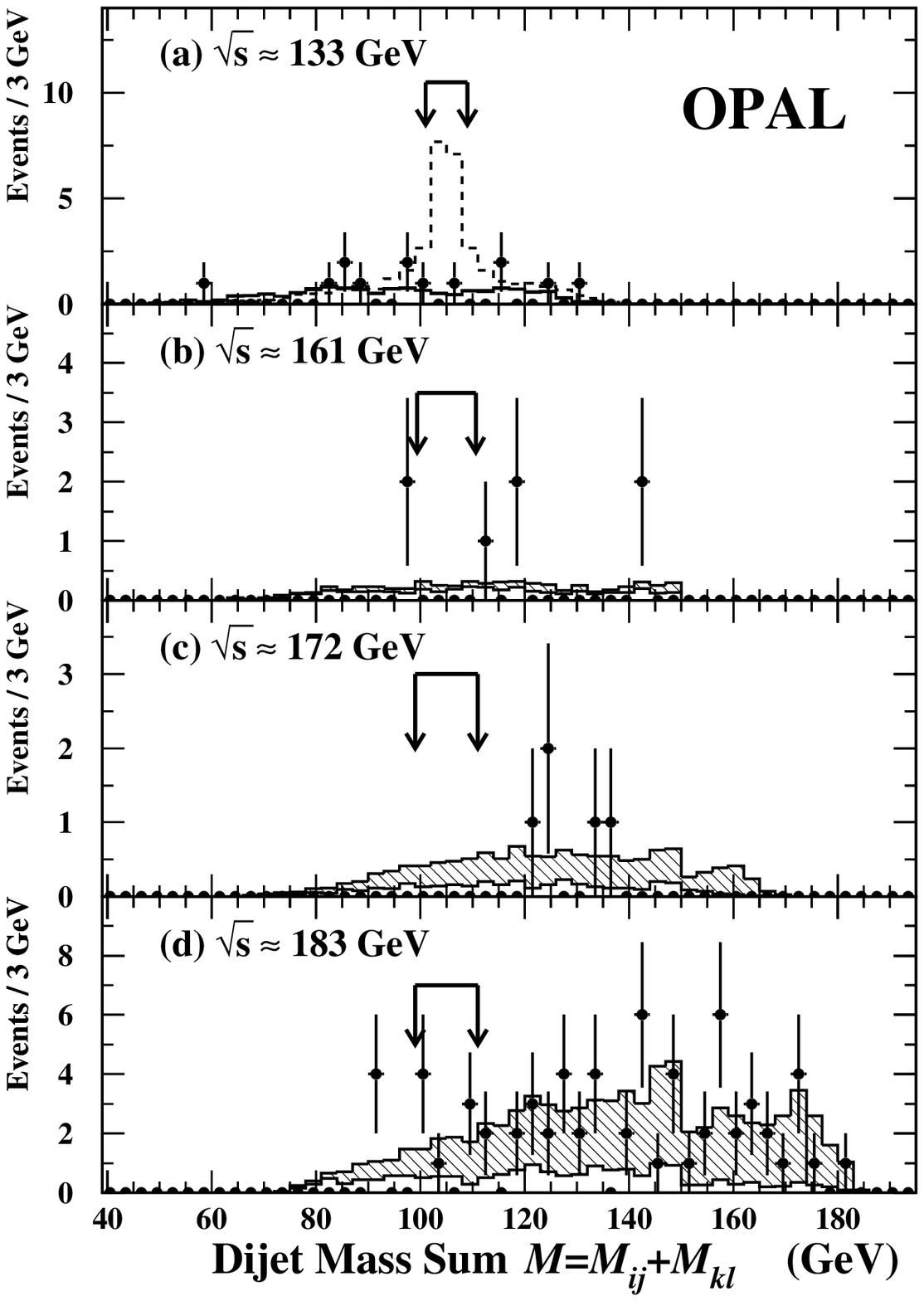}
\caption[]{{\sl
The dijet mass sum for the combination with the smallest $\Delta M$
after the W-pair veto in OPAL's emulation of the ALEPH analysis
shown separately for the different centre-of-mass energies.
Data are shown by the points and
Standard Model backgrounds by the histograms.  
The hatched component of the background
histograms denotes Standard Model four-fermion processes, 
while the unhatched component denotes
\Zqq. The mass windows containing the region of interest are
indicated by the arrows.  The dashed histogram in (a)
illustrates a signal (plus background) that could be 
expected due to $\mathrm{h}^0 \mathrm{A}^0$ with both decaying to
pairs of b-quark jets, and 
$M_{\mathrm h^0}=M_{\mathrm A^0}=52.5$~GeV, normalized to the 
excess observed by ALEPH at 
$E_{\mathrm{cm}} \approx 133$~GeV~\cite{bib-ALEPH4jet}.
}}
\label{fig:allenergy}
\end{center}
\end{figure}

The dijet mass sum for the combination with the smallest
dijet mass difference is shown in Figure~\ref{fig:allenergy},
separately for the data samples at the four different values of $\sqrt{s}$,
after the W-pair veto.  The expected Standard Model background distribution
is shown with the data for each case. 
No significant excess is
observed at any of the centre-of-mass energies.

The sensitivity of the analysis to a peak at a particular dijet
mass sum depends on the resolution and may be affected
by energy scale biases.  The resolution was investigated using
the HZHA event generator~\cite{hzha} to model the process
${\rm e}^+ {\rm e}^-\!\rightarrow \mathrm{h}^0\mathrm{A}^0$.
The masses of both the h$^0$ and the A$^0$ are
taken to be equal to 55~GeV with negligible
width.  The resolutions found for the reconstructed dijet mass sum $M$
for the combination with the smallest $\Delta M$ are 
$\sigma_M = 2.0$, 2.8,  3.0, and 3.0~GeV for
$\sqrt{s}=133$~GeV, 161~GeV, 172~GeV, and 183~GeV, respectively.
These resolutions do not include effects of significant non-Gaussian
tails which arise from wrong jet-pair combinations,
where the correct jet-pair combination
is when each of the two jets comes from the decay of the same particle.
For example, at $\sqrt{s}=133$~GeV, 38\% of the events fall into these tails. 
This definition of the resolution and tails is the same as that used in the
ALEPH publication~\cite{bib-ALEPH4jet}
and the resolution values found are similar to those of ALEPH.
The degradation of mass resolution with increasing energy arises
from the scaling of the jet energies to the beam energy and also from
the energy dependence of the detector resolution.

Studies of the $\mathrm{h}^0\mathrm{A}^0$ signal Monte Carlo with
samples generated with input masses adding to 110~GeV
at each centre-of-mass energy show that
the reconstructed $M$ distributions have peaks at masses
consistent with this input value
within their errors of approximately
0.5~GeV.  Studies of events from radiative returns to the Z$^0$,
$\mathrm{q\bar{q}\gamma}$, were also used to check that the Z$^0$ peak
is well simulated in position and shape, further
indicating that there is no
significant bias in $M$ 
or degradation in resolution
inherent to the selection or the mass
reconstruction procedure.

For the 133~GeV signal Monte Carlo,
if the reconstructed dijet mass sum for the jet-pair combination
having the smallest $\Delta M$ is required to be within 2$\sigma_M$
(4.0~GeV) of the generated mass sum, the efficiency obtained is
26.6\%, which is
60\% of the efficiency obtained without the requirement on $M$.
The efficiencies and expected backgrounds both before and after the mass 
window cut
are similar to those obtained by ALEPH~\cite{bib-ALEPH4jet} so
that for the same integrated luminosity, the observed number of events
can be directly compared to the number observed in Reference~\cite{bib-ALEPH4jet}.
At higher centre-of-mass energies, 
accepting events only in a mass window of width
$\pm 2\sigma_M$ results in efficiencies of 18--21\%  which is
66--69\% of the efficiency before the mass window requirement.

To search for an excess of four-jet events with dijet mass sums near
105~GeV as motivated by Reference~\cite{bib-ALEPH4jet},
events satisfying $|M - 105\ {\rm GeV}| < 2\sigma_M$
for the combination with the smallest $\Delta M$ were counted and the
Standard Model backgrounds were estimated.  
These mass windows are shown in Figure~\ref{fig:allenergy} and
the results of the
searches within these mass regions are given in
Table~\ref{table:masswinevent} both before and after the W-pair veto,
when applicable.  No significant excess is seen in any sample. 
Combining data from all centre-of-mass energies, nine events are observed
while $11.5 \pm 0.4$ are expected from Standard Model processes.

\begin{table}[tbh]
\begin{center}
\begin{tabular}{|c||c|c|c||c|c|c|} \cline{2-7}
 \multicolumn{1}{r|}{ } & \multicolumn{3}{c||}{Without W-Pair Veto}
                        & \multicolumn{3}{c|}{With W-Pair Veto} \\ \hline
 Data Sample & Observed & Expected & Sig. eff., $\epsilon_{\mathrm{hA}}$   
             & Observed & Expected & Sig. eff., $\epsilon_{\mathrm{hA}}$ \\ 
\hline\hline
 133~GeV & 1 & $1.7 \pm 0.2$ & 26.6\% & 1   & $1.7 \pm 0.2$ & 26.6\% \\ 
 161~GeV & 1 & $1.5 \pm 0.1$ & 26.2\% & 0   & $1.0 \pm 0.1$ & 18.4\% \\ 
 172~GeV & 3 & $2.9 \pm 0.1$ & 28.3\% & 0   & $1.8 \pm 0.1$ & 20.7\%  \\ 
 183~GeV &13 &$11.5 \pm 0.4$ & 26.0\% & 8   & $7.0 \pm 0.3$ & 18.3\% \\ \hline
 Total   &18 &$17.7 \pm 0.5$ & ---    & 9   &$11.5 \pm 0.4$ & --- \\ \hline
\end{tabular}
\caption{{\sl Observed event count and expected Standard Model background
for selected events close to 105~GeV,
for the combination with the smallest $\Delta M$, before and after
the W-pair veto.  No W-pair veto is
applied for the 133~GeV data. The mass window is chosen to allow events
that are within $\pm 2\sigma_M$ of 105~GeV to be included, where 
$\sigma_M$ is the
expected experimental resolution on $M$ as given in the text.
Signal efficiencies apply to h$^0$A$^0$ production with 
$M_{\mathrm h^0}=M_{\mathrm A^0}=55$ GeV.}}
\label{table:masswinevent}
\end{center}
\end{table}

\begin{figure}[htb]
\begin{center}
\epsfxsize=16cm
\epsffile{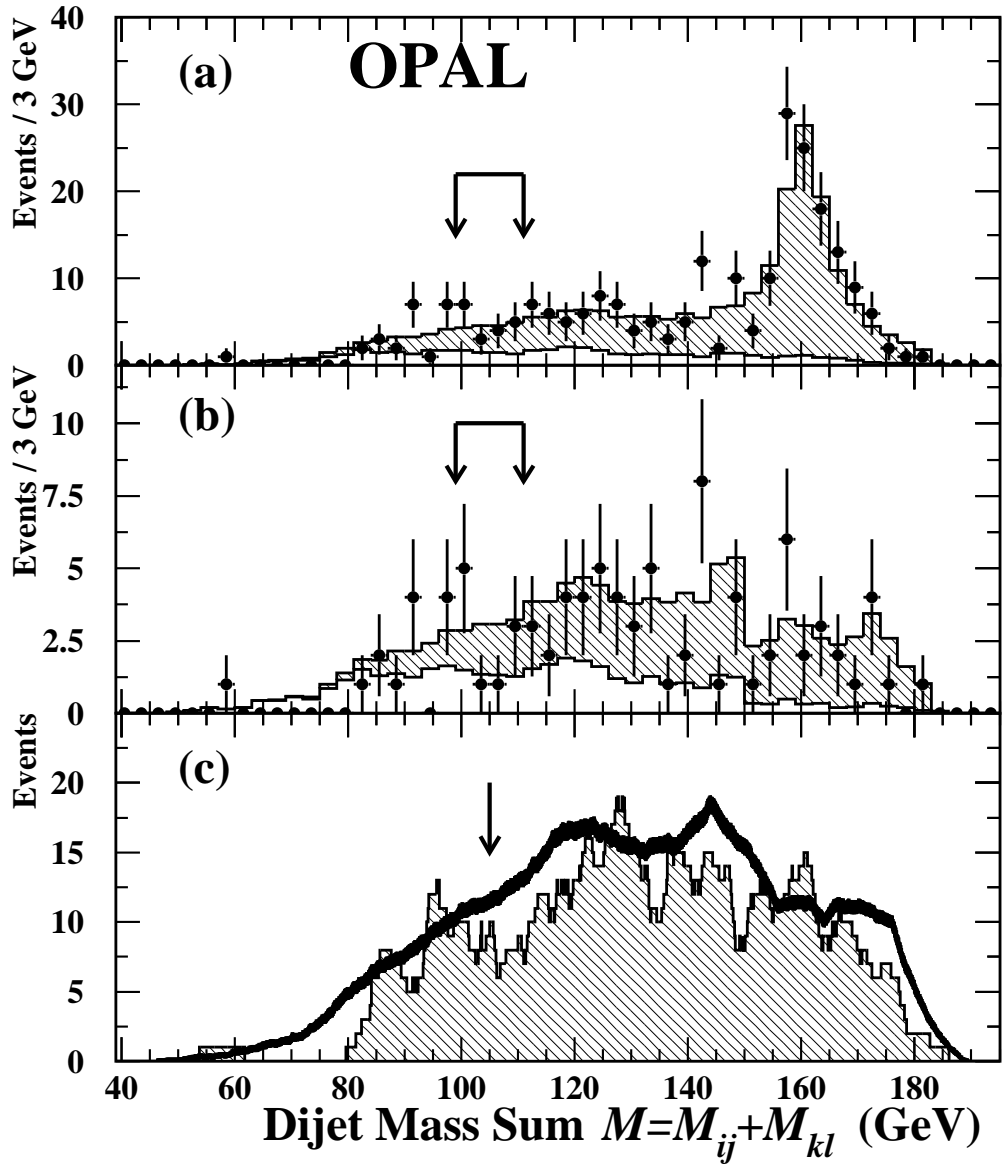}
\caption[]{{\sl 
The dijet mass sum in OPAL's emulation of the ALEPH analysis
for the combined 130--184~GeV samples.
Plots~(a) and~(b) show the distribution of the dijet mass
sum before and after the W-pair veto, respectively. 
Data are shown by the points and
Standard Model backgrounds by the histograms.  
The hatched component of the background
histograms denotes Standard Model four-fermion processes
and the unhatched component denotes
\Zqq.  The mass window around 105~GeV whose width accommodates
the resolution at $\sqrt{s}=183$~GeV, is shown with the
arrows.  Plot~(c) shows the
sliding mass window scan for the same analysis after the
W-pair veto.  The hatched histograms show the total number of data events,
and the solid line shows the Standard Model expectation; the
line width indicates the Monte Carlo statistical error.  An
arrow is drawn at 105~GeV.}}
\label{fig:wvetonowveto}
\end{center}
\end{figure}

Figure~\ref{fig:wvetonowveto} shows the distribution of
the dijet mass sum for the jet pairing with the smallest $\Delta M$
for all running periods combined, with and without the W-pair veto.
The data agree well with the Standard Model
background simulation and no excess is observed
in the region $99.0 < M < 111.0$~GeV, where the width has
been chosen to accommodate the resolution at the highest energy.
A clear peak may be seen at twice the W mass in the sample before
the W-pair veto.

To test for
a peak in the dijet mass sum distribution
for arbitrary mass $M$ and independent of histogram binning,
the positions of the mass windows were scanned over the
full range of $M$.
The results are shown in Figure~\ref{fig:wvetonowveto}(c) for
the combined data samples.
The figure displays the event counts
within windows of fixed width but whose centres are adjusted
in steps of 50~MeV. 
The width of the mass window is $\pm 4.0$, $\pm 5.6$, $\pm 6.0$~GeV,
and $\pm 6.0$~GeV
for the 133, 161, 172, and 183~GeV data samples, respectively, to
reflect the resolution.
The contents of nearby bins in these scans
have high statistical correlations.
No significant excess is observed in the mass
window scan at any value of the dijet mass sum.  In
particular, no choice of binning produces a peak near 105~GeV.

\begin{figure}[htb]
\begin{center}
\epsfxsize=16cm
\epsffile{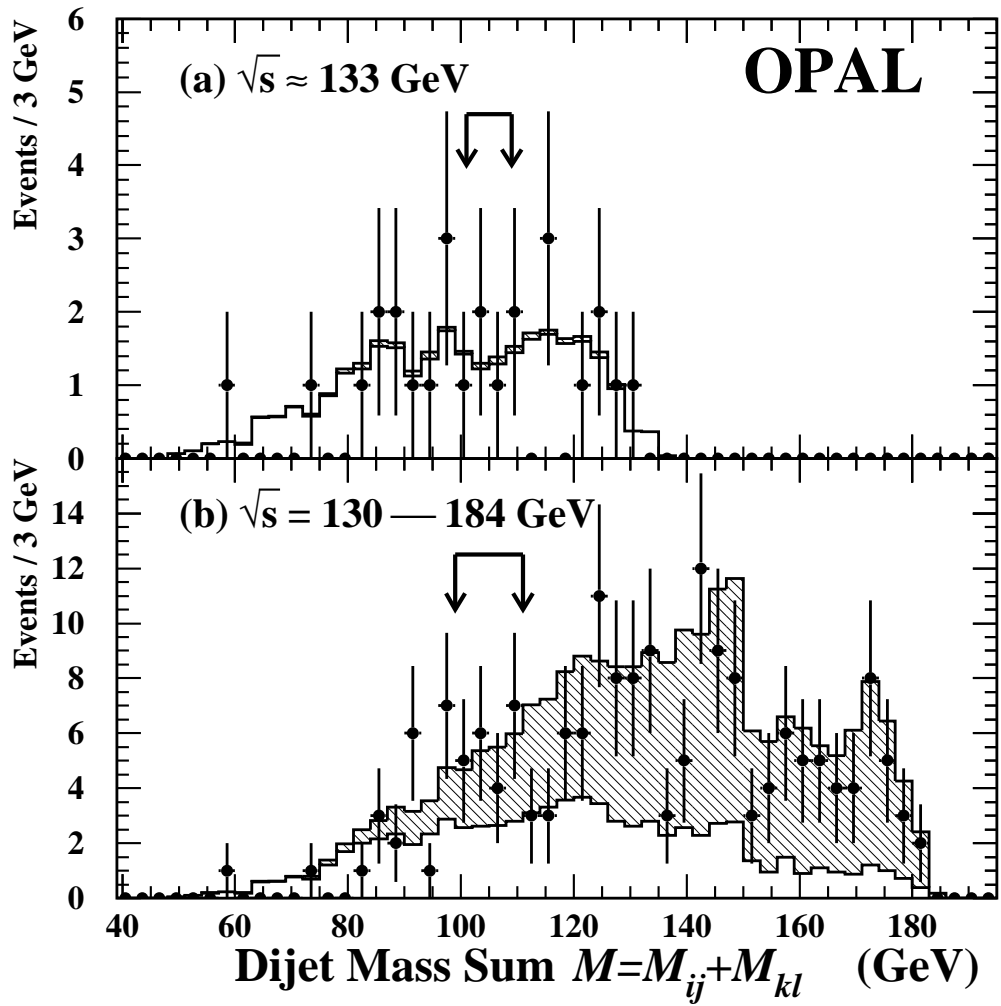}
\caption[]{\sl
The dijet mass sum in OPAL's emulation of the ALEPH analysis
for both the combination with the smallest $\Delta M$
and the combination with the second-smallest $\Delta M$ for
(a) the 133~GeV data sample
and (b) the combined 130--184~GeV samples.
The points and histograms are as in Figure~\ref{fig:allenergy}.
}
\label{fig:seccombo}
\end{center}
\end{figure}

To check for a possible signal in the fraction of events with
wrong jet-pair combinations,
the dijet mass sum for the jet pairing
with the second-smallest $\Delta M$ was also considered.
If the mass difference
of a pair of objects produced together were 20 GeV, the correct
jet pairing would yield the smallest $\Delta M$ 
for roughly half of the signal,
and the second-smallest $\Delta M$ for most of the remainder.
In the ALEPH analysis~\cite{bib-ALEPH4jet}, including the second
combination to the dijet mass sum distribution 
resulted in three additional events within the mass window
with an additional 1.2 events expected from Standard Model processes.
Figure~\ref{fig:seccombo} shows the effect of adding the dijet mass sum
distributions for the smallest and 
second-smallest $\Delta M$ combinations for different
centre-of-mass energies.
The distributions agree well with the Standard Model prediction 
and no peak arises when the second combination
is included.

\subsection{OPAL-Specific Analysis}
\label{secopalspecific}

In the above analysis, an emulation of the ALEPH selection criteria was 
applied to the OPAL data 
to test for the presence of events of the type observed by ALEPH.
The selection described below is an
OPAL-specific analysis in which the sensitivity has been maximised for
detecting a possible signal for the process $\mathrm{e^+e^-} \ra \mathrm{XY}$ 
in the form of an excess
of events with similar mass sums $M=M_{\mathrm X} + M_{\mathrm Y}$
in the four-jet topology.  The analysis is designed to retain
sensitivity even when the mass difference 
$\Delta M=|M_{\mathrm X} - M_{\mathrm Y}|$ is as large as 30~GeV.
Efficiencies and backgrounds are estimated
for different values of $M$ and $\Delta M$.

The cuts are designed to be as insensitive as possible to the 
flavours of the final state quarks.
Although the methods employed at each of the centre-of-mass energies
are similar, the optimal cut values in most cases depend on~\sqrts.
\begin{itemize}
\item[1.]
      The events must pass the hadronic final state requirement
      of cut 1 in Section~\ref{secalephemul}.
\item[2.]
      The effective centre-of-mass energy after initial-state radiation,
      \sqrtsp, calculated using the method described in Reference~\cite{sprim}, 
      has to be at least 0.87\sqrts.  The measured visible mass,
      $M_{\rm vis}$, is required to be  between \sqrts$-40$~GeV and 
      \sqrts$+30$~GeV  
      at 133~GeV, between 100 and 200~GeV at \sqrts$=161$~GeV, 
      between 110 and 210~GeV at \sqrts$=172$~GeV and 
      between 120 and 220 GeV at \sqrts$=183$~GeV.
      
\item[3.]
      The charged particles and calorimeter clusters are grouped
      into four jets using
      the Durham algorithm~\cite{bib:durham}. The jet resolution parameter,
      $y_{34}$, at which the number of jets changes from 
      three to four, is required to
      be larger than 0.007 at $\sqrt{s}=133$~GeV, and
      larger than 0.005 at $\sqrt{s}=$161--183~GeV.
      To discriminate against poorly reconstructed events, a kinematic fit
      imposing energy and momentum conservation is required
      to yield a $\chi^2$ probability larger than 0.01. 
      Each of the four jets is
      required to contain at least two tracks at 133~GeV
      and at least one track at higher energies.
      These kinematically constrained jets are used in the 
      subsequent calculation of dijet masses.

\item[4.]
      In the case of the 161--183~GeV data, 
      the background from \ee\ra \Zg\ is further reduced by eliminating those
      events where one of the four jets has properties compatible
      with those of a radiative photon, namely that it has
      exactly one electromagnetic
      cluster, not more than two tracks (possibly from a photon conversion), 
      and energy between
      45 and 65~GeV at \sqrts$=161$~GeV, between 52 and 72~GeV at \sqrts$=172$~GeV
      and between 60 and 80 GeV at \sqrts$=183$~GeV.

\item[5.]
      The  polar angle of the thrust axis,
      $\theta_{\mathrm{thr}}$, is required to satisfy 
      $|\cos\theta_{\mathrm{thr}}| < 0.9$
      at 133~GeV and $|\cos\theta_{\mathrm{thr}}| < 0.8$ at 161--183 GeV.
      
\item[6.]
      To reduce background from
      $\mathrm{q\bar{q}}$ events,
      the event shape parameter $C$ \cite{cpar}, which 
      ranges between 0 and 1 and is
      0 for a perfect 2-jet event, 
      is required to be larger than 
      0.7  at  $\sqrt{s}=$133~GeV
      and larger than 0.6 at higher energies.

\item[7.]
      To ensure well-separated jets for better kinematic fits,
      the angle between any two jets is required to exceed 0.8 radians
      for 161--183 GeV data.

\item[8.]
      Above the \WW\ threshold, explicit 
      vetoes against the process  \ee\ra \WW\ are applied.

      At $\sqrt{s}=$161~GeV, the two \W~bosons are produced
      with only a small boost.
      The two jets having the largest opening angle are assigned
      to one of the \W~bosons and the two remaining jets to the
      other. An event is rejected if both jet pairs have an invariant
      mass between 75~GeV and 90~GeV. 
      
      At $\sqrt{s}=$172 and 183~GeV, a more sophisticated veto is applied. 
      The four jets are combined into pairs, and for all three
      combinations the event is refitted constraining the total
      energy to $\sqrt{s}$ and the total momentum
      to zero, and also constraining the masses of the
      two jet pairs to be equal (five constraints). 
      From the three combinations, the one yielding the largest
      $\chi^2$ fit probability is considered.  If the jet pair
      mass from the fit exceeds 75~GeV and the fit
      probability is at least 0.01, the event is rejected.

\item[9.]
To achieve good sensitivity for all \Dm\ less than 30~GeV, we use
two separate mass selections, one relevant for unequal masses 
and one for equal masses. 
In both selections, when searching for a signal with a
hypothetical sum of masses, $M_0$, the range
$M_0 \pm 2\sigma_M$ is used, 
where $\sigma_M$ is 2.0~GeV at $\sqrt{s}=$133~GeV and
3.0~GeV for higher energies.
For unequal masses (\Dm\ $>5$~GeV), 
the event is selected if either the jet association with
the smallest mass difference or the one with the second smallest 
mass difference has
a mass sum \m\ in the range $M_0 \pm 2\sigma_M$.
For nearly equal masses (\Dm\ $<5$~GeV), better sensitivity 
is obtained when considering only the jet association with
the smallest mass difference. 
The resolution $\sigma_M$ varies only slowly with $M$ and \Dm.
\end{itemize}

Table \ref{OPAL_exp} presents the number of observed events
and the Standard Model expectations before and after the \WW\ veto (cut 8).
The numbers of observed events are consistent with the
Standard Model expectations at all centre-of-mass energies
both before and after the mass selection.

\begin{table}[tbh]
\begin{center}
\begin{tabular}{|c||c|c||c|c||c|c|} \cline{2-7}
 \multicolumn{1}{r|}{ } & \multicolumn{2}{c||}{Without W-Pair Veto}
                        & \multicolumn{2}{c||}{With W-Pair Veto} 
                        & \multicolumn{2}{c|}{After Mass Selection}\\ \hline
 Data Sample & Observed & Expected  
             & Observed & Expected 
             & Observed & Expected \\ 
\hline\hline
 133~GeV & 18   & $17.0  \pm 0.6$ & 18  & $17.0  \pm 0.6$ & 4 & $3.1 \pm 0.3$ \\ 
 161~GeV & 11   & $15.8  \pm 0.3$ & 8   & $13.6  \pm 0.3$ & 2 & $2.7 \pm 0.1$ \\ 
 172~GeV & 36   & $33.8  \pm 0.3$ & 21  & $16.2  \pm 0.2$ & 4 & $2.9 \pm 0.1$ \\ 
 183~GeV & 190  & $210.1 \pm 1.2$ & 70  & $81.6  \pm 0.8$ & 6 & $8.9 \pm 0.3$ \\ \hline
 Total   & 255  & $276.7 \pm 1.4$ & 117 & $128.4 \pm 1.1$ &16 & $17.6\pm 0.4$ \\ \hline
\end{tabular}
\caption[sig]{\label{OPAL_exp} {\sl Number of observed and
expected Standard Model background events before 
and after the W-pair veto and after adding the mass selection 
(cut 9)
centred at 105~GeV for the smallest $\Delta M$ combination. 
The quoted errors are statistical.  No W-Pair veto has been applied
to the 133~GeV data.}}
\end{center}
\end{table}

Table~\ref{OPAL_eff} shows the signal efficiencies 
for various combinations of (\mX,\mY) together with 
the predicted background and the numbers of
observed events after all cuts.

\begin{figure}[htb]
\begin{center}
\epsfxsize=16cm
\epsffile{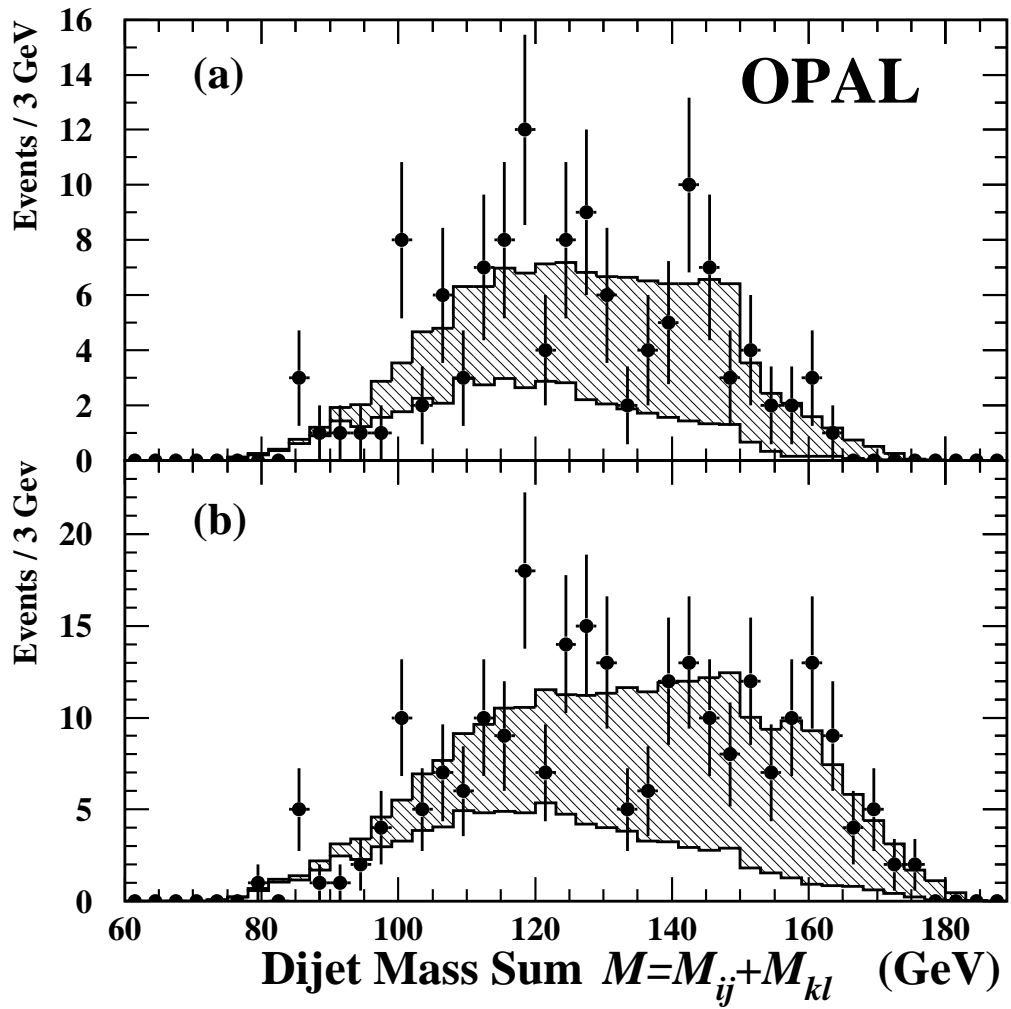}
\caption[]{\sl
Distributions of \m\ in the OPAL-specific analysis for the
combined 130--184~GeV samples after all selection
requirements except the mass selection (cut 9),
(a) for the jet combination with the smallest \Dm\ and 
(b) for the jet combinations with the smallest and second-smallest \Dm.
The points and histograms are as in Figure~\ref{fig:allenergy}.
}
\label{OPAL_mass}
\end{center}
\end{figure}

Figure~\ref{OPAL_mass} shows the
distributions of \m\ for the jet associations with the smallest \Dm, 
and for the jet association with the smallest and second-smallest \Dm,
summed over all centre-of-mass energies.
Globally, the distributions show consistency between the data 
and the Standard Model background prediction. In
particular, there is no excess in the vicinity of \m\ $\approx 105$~GeV.
The overlap of the OPAL-specific analysis and the OPAL emulation of the ALEPH
selection has been evaluated in a typical 
Monte Carlo four-jet signal sample at $\sqrt{s}= 133$~GeV
with $M_{\mathrm h^0}=M_{\mathrm A^0}=55$ GeV. In this sample,
59\% of the events selected by the OPAL emulation of the ALEPH analysis
are also selected by the OPAL-specific analysis.

\subsection{Systematic Errors}

At $\sqrt{s}= 133$~GeV, since the efficiencies and expected backgrounds 
for the OPAL emulation of the ALEPH signal
are similar to those obtained by ALEPH~\cite{bib-ALEPH4jet}, it is not
necessary to consider systematic effects in detail if only numbers of
observed events are compared.  However, to calculate limits on 
cross-sections, systematic errors on efficiencies and backgrounds are estimated.

To emulate the ALEPH analysis, the charged multiplicity
requirement (cut 9, Section~4.1) on all combinations of two jets was necessary.  
Since the
aim of the emulation analysis is to compare directly the OPAL result
with the ALEPH observation, and ALEPH's cross-section estimate was
made assuming a model of four b-jets, we also assume this model.
Varying the mean charged multiplicity in b-hadron decays by its
measurement uncertainty~\cite{bmult} 
results in an estimated systematic error of 
12\% on signal detection efficiencies due to this effect.  
Including additional uncertainties in the 
modelling of the
cut variables, energy scales, mass resolutions and limited
Monte Carlo statistics results in an estimated total systematic
error of 13\%.

\begin{table}[htb]
\begin{center}
\begin{tabular}{|c||c|c|c||c|c|c|} \cline{2-7}
\multicolumn{1}{c|}{ } & \multicolumn{3}{c||}{133 GeV} & \multicolumn{3}{c|}{161 GeV} \\ \hline
(\mX,\mY)        & Effic.     & Backgd. & Data   & Effic.     & Backgd. & Data \\  
(GeV)            & (\%)       &         &        & (\%)       &         &      \\
\hline\hline
(50,50)& $30.6\pm 1.5$ & $3.2\pm 0.3$ & 3&  $38.0\pm 2.2$ & $2.4\pm 0.1$ &     3 \\        
(40,60)& $29.6\pm 2.0$ & $5.6\pm 0.4$ & 6&  $35.8\pm 2.1$ & $4.2\pm 0.2$ &     3 \\ \hline
(55,55)& $29.0\pm 2.0$ & $3.1\pm 0.3$ & 4&  $37.4\pm 1.0$ & $3.0\pm 0.1$ &     1 \\        
(50,60)& $38.4\pm 1.5$ & $5.6\pm 0.4$ & 7&  $42.0\pm 2.2$ & $5.0\pm 0.2$ &     1 \\       
(40,70)& $23.0\pm 1.9$ & $5.6\pm 0.4$ & 7&  $34.8\pm 2.1$ & $5.0\pm 0.2$ &     1 \\ \hline  
(60,60)& $26.2\pm 1.4$ & $3.2\pm 0.3$ & 3&  $39.0\pm 2.2$ & $2.8\pm 0.1$ &     0 \\         
(50,70)& $30.2\pm 2.1$ & $5.6\pm 0.4$ & 4&  $41.2\pm 2.2$ & $4.8\pm 0.2$ &     1 \\ \hline  
(60,70)& $24.2\pm 1.9$ & $2.2\pm 0.2$ & 4&  $34.2\pm 2.1$ & $4.7\pm 0.2$ &     2 \\         
(50,80)& $13.1\pm 1.1$ & $2.2\pm 0.2$ & 4&  $32.8\pm 2.1$ & $4.7\pm 0.2$ &     2 \\ \hline 
(70,70)& ---  & --- &              --- &  $26.6\pm 2.0$ & $2.0\pm 0.1$ &  2 \\
(60,80)& ---  & --- &              --- &  $33.8\pm 2.1$ & $4.4\pm 0.2$ &  3 \\
\hline
\multicolumn{7}{c}{ } \\
\multicolumn{7}{c}{ } \\ \cline{2-7}
\multicolumn{1}{c|}{ }   & \multicolumn{3}{c||}{172 GeV} & \multicolumn{3}{c|}{183 GeV} \\ \hline
(\mX,\mY)        & Effic.     & Backgd. & Data   & Effic.     & Backgd. & Data \\  
(GeV)            & (\%)       &         &        & (\%)       &         &      \\
\hline\hline
(50,50)&$31.0\pm 1.5$&$2.5\pm 0.1$& 4 & $19.4\pm 1.8$  & $4.8\pm 0.2$  & 4 \\
(40,60)&$24.6\pm 1.9$&$3.2\pm 0.1$& 5 & $16.3\pm 1.6$  & $5.6\pm 0.2$  & 6 \\ \hline
(55,55)&$34.4\pm 1.0$&$3.6\pm 0.2$& 5 & $31.6\pm 2.1$  & $12.4\pm 0.3$ & 9 \\
(50,60)&$32.6\pm 1.5$&$5.1\pm 0.2$& 6 & $22.2\pm 1.9$  & $16.2\pm 0.3$ & 12\\
(40,70)&$23.0\pm 1.3$&$5.1\pm 0.2$& 6 & $18.0\pm 1.7$  & $16.2\pm 0.3$ & 12\\ \hline
(60,60)&$33.0\pm 1.5$&$3.7\pm 0.2$& 9 & $32.6\pm 2.1$  & $17.1\pm 0.4$ & 20\\
(50,70)&$34.4\pm 2.1$&$5.5\pm 0.2$& 11& $23.9\pm 1.9$  & $23.5\pm 0.4$ & 30\\ \hline
(60,70)&$33.6\pm 1.5$&$5.5\pm 0.2$& 6 & $32.2\pm 2.1$  & $28.2\pm 0.4$ & 24\\
(50,80)&$28.8\pm 1.4$&$5.5\pm 0.2$& 6 & $23.2\pm 1.9$  & $28.2\pm 0.4$ & 24\\ \hline
(70,70)&$29.9\pm 1.4$&$3.3\pm 0.1$& 1 & $31.2\pm 2.1$  & $20.7\pm 0.4$ & 20\\
(60,80)&$31.0\pm 2.1$&$6.0\pm 0.2$& 4 & $27.7\pm 2.0$  & $32.3\pm 0.5$ & 28\\
\hline
\end{tabular}
\end{center}
\caption[sig]{\label{OPAL_eff}  {\sl Signal detection efficiencies,
numbers of expected background events and number of observed data events,
for various mass combinations in the OPAL-specific analysis, after
the mass selection, cut 9 of section~4.2. The quoted errors are statistical.}
}
\end{table}

In the OPAL-specific analysis, 
the signal detection efficiencies are subject to a systematic 
error of 9\%,
which
includes an allowance for the final state to contain any composition
of quark flavours and uncertainties in modelling heavy hadron decays,
the uncertainty on the simulation of the  decay
with regards to fragmentation and hadronization,
the modelling of the cut variables,
and the limited Monte Carlo statistics.

The total relative uncertainty on the residual
background is 20\% for the OPAL emulation of the
ALEPH analysis, and 13\% for the OPAL-specific analysis.  
These errors include the uncertainty on the modelling of the
hadronization process, on the prediction of the four-jet rate,
W-pair cross-section,
 and the 
modelling of
the cut variables.  The error due to the limited Monte Carlo statistics
is added in quadrature to this uncertainty.
The systematic errors on the luminosity measurements range from
0.5\% to 1.6\%.

\section{Cross-Section Upper Limits}

In the OPAL emulation of the ALEPH analysis, the number of observed
events 
can be compared directly to the ALEPH
observation~\cite{bib-ALEPH4jet} because both the observed background
rate and the estimated efficiency are nearly identical
to those obtained by ALEPH.
From the number of observed and
expected events in the dijet mass sum window of $105 \pm 4$~GeV at 
$\sqrt{s} \approx 133$~GeV, we set
a 95\% confidence level (CL) upper limit
of 2.1 events that could be attributed to 
additional cross-section from new physics 
when scaled to the integrated
luminosity of the 1995 ALEPH result.
This can be compared 
to ALEPH's observation in 1995 of nine events with a Standard Model 
expectation of 0.8 events.
To calculate the probability that the OPAL observation is consistent
with the ALEPH observation in the presence of a possible signal,
the product is formed of
the Poisson probability $p_1$ that at least nine events were observed
in ALEPH and $p_2$ that no more than one event was observed in OPAL, given
the Standard Model backgrounds and assuming the presence of a signal
scaled by the integrated luminosity.  
The probability of an outcome no more likely than                            
that observed in the data, i.e., the sum of Poisson probabilities            
of possible outcomes less than or equal to $p_1 p_2$,                        
is found to be 2.6$\times 10^{-4}$, where the hypothesized signal           
cross-section has been chosen to maximize this probability.  
 
Assuming  production of new 
particles X and Y
subsequently decaying to a final state of four b-jets
to determine efficiencies, cross-section upper limits
are set using the OPAL emulation of the ALEPH analysis.
From the number of observed and
expected events in the dijet mass sum window as above at 
$\sqrt{s} \approx 133$~GeV, 
 a 95\% confidence level (CL) upper limit
of 1.4~pb is determined for the
production cross-section at the dijet mass sum of 105~GeV.  
Systematic
uncertainties of efficiency, background and luminosity 
are taken into account
using the procedure outlined in Reference~\cite{cousins}.

To combine data from different centre-of-mass energies, we consider
two different functions for the energy dependence of the cross-section
of a hypothetical signal.
It is first assumed that the cross-section varies as
$\beta (3 - \beta^2)/s$, typical for pair-production of spin-1/2 particles,
where $\beta$ is taken as the average velocity of the particles
in the laboratory frame~\cite{collider}.
Taking from Table~\ref{table:masswinevent} the total number of observed
and expected events in resolution-dependent
mass windows around 105~GeV,
an upper limit on the production cross-section at 133~GeV
of 0.58~pb at 95\% CL is found.  
Secondly, under the hypothesis that the signal cross-section
varies as $\beta^3/s$, typical for the 
production of scalar particles~\cite{collider}, the upper limit on the 
cross-section at 133~GeV is computed to be 0.31~pb at 95\% CL.
These limits can be compared to ALEPH's estimated cross-section of
$3.1 \pm 1.7$~pb~\cite{bib-ALEPH4jet} from their total number of
excess events.

\begin{figure}[htb]
\begin{center}
\epsfxsize=16cm
\epsffile{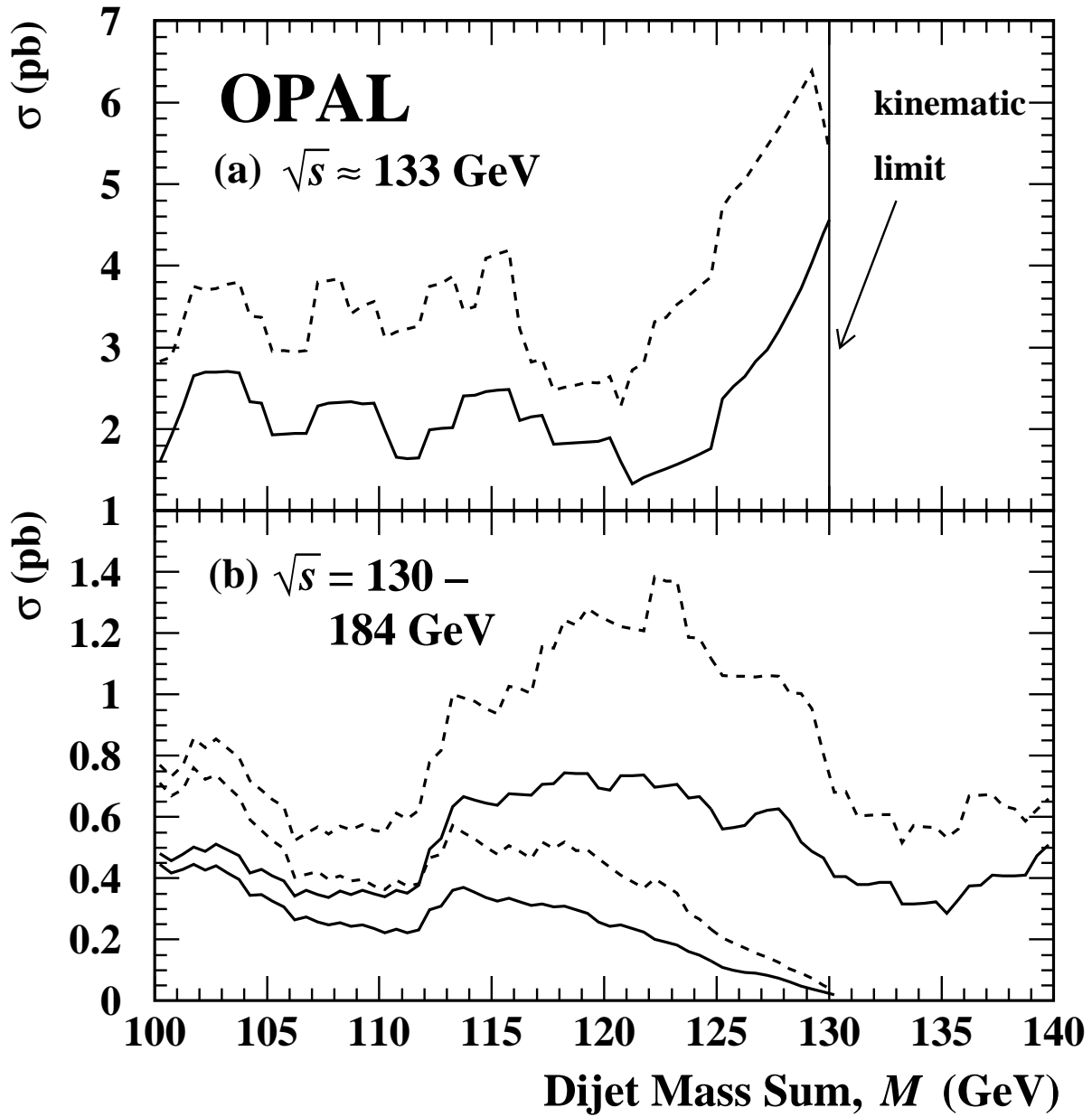}
\caption[]{\sl The 95\% CL upper limits obtained with the
OPAL-specific analysis on the production cross-section
of a possible signal as a function of \m\ 
for \Dm\ close to 0 (solid lines) and for \Dm\ $<30$ GeV (dashed lines).
Plot~(a) shows the limits computed using the data collected
at \sqrts$\approx$133~GeV; plot~(b) shows the limits using
the combined data from $\sqrt{s} = 130$--184~GeV
assuming a cross-section that varies as $\beta^3/s$ scaled to
$\sqrt{s} = 133$~GeV (lines that end near \m=130~GeV)
and  $\sqrt{s} = 183$~GeV (lines that extend to larger \m).}
\label{OPAL_lim}
\end{center}
\end{figure}

The OPAL-specific analysis described in section~4.2
is used to obtain upper limits for the cross-section of a possible
signal process $\mathrm{e^+e^-} \ra \mathrm{XY} \ra$~four jets, in 
the presence of background from Standard Model processes,
using Poisson statistics and incorporating systematic uncertainties
as described in Reference~\cite{cousins}.
The process
$\mathrm{e^+e^-} \ra \mathrm{h^0 A^0}$ was used to
model the signal detection efficiencies.
The resulting 95\%~CL upper limits, as function of the
mass sum $M (\equiv M_{\mathrm{X}} + M_{\mathrm{Y}})$,
are shown in Figure~\ref{OPAL_lim}, 
for $\Delta M$ close to zero and $\Delta M = 30$~GeV. 
A mass window of $M \pm 2\sigma_M$ is scanned across the distribution
of the dijet mass sum in small steps.
To account for a possible discrepancy between the mass scale of the
data and the Monte Carlo in a conservative manner,
the mass window is displaced by $\pm 0.5$~GeV at each scan point.  
The largest data count in any of
the three windows 
including the nominal one and the smallest background estimation in any of 
the three
windows are used to compute the limit.
When results at different centre-of-mass energies are combined,
the hypothetical production cross-section 
is assumed to vary as $\beta^3/s$.
The cross-section limits are presented 
separately for the 133~GeV data, and for all
data (130--184~GeV) combined.  Limits on the cross-section
from the combined data sample are computed both at $\sqrt{s} = 133$~GeV
and at $\sqrt{s} = 183$~GeV.
These results are independent of the
flavour of the quarks from the decay of the hypothesized particles and
are valid for 
X and Y being scalars produced predominantly
by an $s$-channel process.

\section{Conclusions}

Following the ALEPH observation of a large excess of four-jet events
with dijet mass sums around 105 GeV at 
$\sqrt{s} \approx 133$~GeV~\cite{bib-ALEPH4jet},
a careful emulation of the ALEPH analysis has been performed using OPAL
data collected from e$^+$e$^-$ collisions at centre-of-mass energies
between 130 and 184~GeV.
  The process 
$\mathrm{e}^+\mathrm{e}^-\rightarrow\mathrm{h}^0\mathrm{A}^0$
was used to estimate the signal detection efficiencies.
The estimated sensitivity, mass resolution, efficiency, and
estimated backgrounds in this analysis
were similar to that of the ALEPH analysis.
No significant excess of four-jet events with dijet
mass sums in the region close to 105 GeV,   
or any other region between 60 and 160 GeV,
has been observed in any of the data samples separately or combined,
and our observations are consistent with Standard Model predictions.
The same conclusions are reached
when an OPAL-specific analysis is employed.
Limits for the cross-section of a hypothetical process
$\mathrm{e^+e^-} \ra \mathrm{XY} \ra$~four jets are given
as a function of the dijet mass sum $M$ and the mass difference $\Delta M$.
The 95\% confidence upper limits obtained in both analyses
for dijet mass sums near
105~GeV are below the excess reported in 1995
by ALEPH~\cite{bib-ALEPH4jet} to a high degree of confidence.
ALEPH has recently analysed~\cite{bib-ALEPHnew} new data at centre-of-mass
energies between 130 and 184~GeV and do not
confirm the previously reported excess.


\par
\vspace*{1.cm}
\section*{Acknowledgements}
\noindent

We particularly wish to thank the SL Division for the efficient operation
of the LEP accelerator at all energies
 and for
their continuing close cooperation with
our experimental group.  We thank our colleagues from CEA, DAPNIA/SPP,
CE-Saclay for their efforts over the years on the time-of-flight and trigger
systems which we continue to use.  In addition to the support staff at our own
institutions we are pleased to acknowledge the  \\
Department of Energy, USA, \\
National Science Foundation, USA, \\
Particle Physics and Astronomy Research Council, UK, \\
Natural Sciences and Engineering Research Council, Canada, \\
Israel Science Foundation, administered by the Israel
Academy of Science and Humanities, \\
Minerva Gesellschaft, \\
Benoziyo Center for High Energy Physics,\\
Japanese Ministry of Education, Science and Culture (the
Monbusho) and a grant under the Monbusho International
Science Research Program,\\
German Israeli Bi-national Science Foundation (GIF), \\
Bundesministerium f\"ur Bildung, Wissenschaft,
Forschung und Technologie, Germany, \\
National Research Council of Canada, \\
Research Corporation, USA,\\
Hungarian Foundation for Scientific Research, OTKA T-016660, 
T023793 and OTKA F-023259.\\

\clearpage
\newpage

\end{document}